\documentstyle[prb,aps,epsf]{revtex}
\draft

\begin{document}

\title{ Spin stripe dimerization in the $t-J$ model.}
\author{O.P. Sushkov}

\address{School of Physics, University of New South Wales,\\
 Sydney 2052, Australia}

\maketitle

\begin{abstract}
In the present work we demonstrate that the spin columnar dimerized
phase (conducting spin stripe liquid) is stable at doping 
$x_{c1} < x < x_{c2}$. At $x < x_{c1}$ the system 
undergoes phase  transition to the Neel state, and at $x > x_{c2}$ 
 it is Normal Fermi liquid. 
At $t/J=3$ the critical concentrations are
$x_{c1}\approx 0.09$ and $x_{c2}\sim 0.36$.
To prove stability of the stripe phase and to calculate critical
concentrations we employ two approaches. The first one is more
technically involved, and it gives more accurate value of the
critical concentration. The approach is based
on the calculation of the magnon Green's function. Imaginary poles in the
Green's function indicate transition to the Neel state.
The second approach consists in direct comparison of ground state
energies of the Neel state and the columnar dimerized state.
Both approaches demonstrate stability of the spin stripe phase and
give close values of the critical concentrations.
\end{abstract}

\section{Introduction}
It is widely believed that the 2D $t-J$ model is relevant to the
low energy physics of high-temperature superconductors.
This is why investigation of this model is of great interest
both for theory and experiment. In spite of great efforts
during more than a decade
there is no full understanding of the phase diagram of the $t-J$ model,
however some facts are well established. At zero doping the model is equivalent
to the Heisenberg model on a square lattice which has long range 
Neel order \cite{O1}. Doping by holes destroys the order.
A simplified picture of noninteracting holes leads to the Neel state
instability with respect to spirals at arbitrary small but finite 
doping \cite{SS}.
However more sophisticated numerical calculations which take into
account renormalization of the hole Green's function under the doping
indicate that the Neel order is stable below some critical hole
concentration $x_{c1}$ \cite{Ig}.
In the Neel phase ($x < x_{c1}$), in all waves except s-wave, 
there is magnon mediated superconducting pairing between holes.
\cite{Fl}.
It is also clear that at very small hopping 
there is phase separation in the model because separation leads to reduction of
the number of destroyed antiferromagnetic links \cite{Emery}

The purpose of the present work is to elucidate spin structure of the 
ground state  at $x > x_{c1}$. The most 
important hint comes from experiment:
indications of stripes in the high-$T_{c}$ materials \cite{stripes}.
Another important hint is a remarkable stability of the spin dimerized phase
in the frustrated $J_1-J_2$ model. The idea of such state for this model
was first formulated by Read and Sachdev \cite{Read}, and was then confirmed by
further work \cite{Gel,Kot}. The stability of
 such a configuration implies that the lattice symmetry is
 spontaneously broken and the ground state is four-fold
 degenerate.
Such a route towards quantum disorder is known rigorously
 to take place in one dimension, where the Lieb-Schultz-Mattis
(LSM) theorem guarantees that a gapped phase always breaks the
 translational symmetry \cite{LSM}.
Some time ago Affleck suggested that the
 LSM theorem  can be extended to higher dimensions, and
 the gapped states of quantum systems necessarily break the discrete 
 symmetries of the lattice \cite{Affleck}.
The example of the $J_1-J_2$ model provides further support for this idea.

There have been several attempts to consider the spin-dimerized phase in a 
doped
Heisenberg antiferromagnet. For this purpose Affleck and Marston \cite{Mar} 
analyzed Hubbard-Heisenberg model in the weak-coupling regime,
Grilli, Castellani and G. Kotliar \cite{Grilli} considered $SU(N)$, 
$N\to \infty$, $t-J$ model, and very recently
Vojta and Sachdev \cite{Voj} considered
$Sp(2N)$, $N\to \infty$,  $t-J$ model.
These works indicated a stability of
the spin-dimerized phase in some region of parameters, providing a very
important guiding line. However relevance of these results to
 "physical regime" of the $t-J$ model remained unclear.
Stability of the spin-dimer order for the $t-J$ model has been demonstrated 
in the paper \cite{dimer}. The only small parameter used in the analysis was
hole concentration with  respect to the half filling. 
In the present work we continue studies in the same direction applying
various techniques.
To be confident in the results we prove stability of the dimer phase
by two independent methods: 1) Calculation of the magnon Green's function
in the dimerized phase (Green Function Method), 2) Comparison of the
ground state energies of the doped Neel state and the dimerized state
(Direct Energy Method). The second approach is very simple physically
and technically. The first approach is more technically involved,
but it allows us to calculate the critical concentration more
precisely.

To incorporate some experimental data we consider
 $t-t^{\prime}-t^{\prime \prime}-J$ model defined by the following
Hamiltonian
\begin{equation}
\label{H}
H=-t\sum_{\langle ij \rangle \sigma} c_{i\sigma}^{\dag}c_{j\sigma}
-t^{\prime}\sum_{\langle ij_1 \rangle \sigma}
c_{i\sigma}^{\dag}c_{j_1\sigma}
-t^{\prime \prime}\sum_{\langle ij_2 \rangle \sigma} 
c_{i\sigma}^{\dag}c_{j_2\sigma}
+ \sum_{\langle ij \rangle \sigma} J_{ij} \left({\bf S}_i{\bf S}_j
-{1\over 4}n_in_j\right).
\end{equation}
 $c_{i \sigma}^{\dag}$ is the  creation operator of an electron with
 spin $\sigma$ $(\sigma =\uparrow, \downarrow)$ at site $i$
 of the two-dimensional square lattice. The $\langle ij \rangle$ represents
nearest neighbor sites, $\langle ij_1 \rangle$ - next nearest neighbor
(diagonal), and $\langle ij_2 \rangle$ represents next next
nearest sites. The spin operator is ${\bf S}_i={1\over 2}
c_{i \alpha}^{\dag} {\bf \sigma}_{\alpha \beta} c_{i \beta}$
and the number density operator is 
$n_i=\sum_{\sigma}c_{i\sigma}^{\dag}c_{i\sigma}$.
The $c_{i\sigma}^{\dag}$ operators act in the Hilbert space with no
double electron occupancy. Antiferromagnetic
interactions $J_{ij}> 0$ are arranged in a stripe pattern shown in Fig. 1:
solid links correspond to $J_{ij}=J_{\perp}=J(1+\delta)$, and
dashed links correspond to $J_{ij}=j=J(1-\delta)$.

\begin{figure}[h]
\vspace{-2pt}
\hspace{-35pt}
\epsfxsize=6cm
\centering\leavevmode\epsfbox{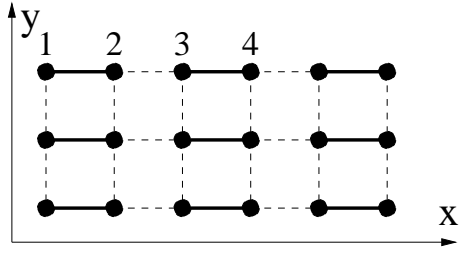}
\vspace{8pt}
\caption{\it {Stripe spin dimerization on square lattice. Solid links
correspond to $J_{\perp}=J(1+\delta)$, and
dashed links correspond to $j=J(1-\delta)$}}
\label{Fig1}
\end{figure}

For real cuprates the antiferromagnetic interaction is isotropic, $\delta=0$.
However for theoretical analysis of the magnon Green's function it is 
convenient to consider nonzero $\delta$ and later set  $\delta \to 0$.
The antiferromagnetic exchange measured in two magnon Raman 
scattering \cite{Tok} is $J=125meV$. Calculation of the
hopping matrix elements has been done by Andersen {\it et al}
\cite{And}. They consider a  two-plane situation, and the effective
matrix elements are slightly different for symmetric and
 antisymmetric combinations of orbitals between planes. After averaging
over these combinations we get: $t=386meV$, $t^{\prime}=-105meV$,
$t^{\prime \prime}=86meV$. Below we set $J=1$, in these units
\begin{equation}
\label{ts}
t=3.1, \ \ \ t^{\prime}=-0.8, \ \ \ t^{\prime \prime}=0.7
\end{equation}
These values are confirmed by the analysis  \cite{Saw} of photoemission 
(PES) data for charge transfer insulator Sr$_2$CuO$_2$Cl$_2$.
In further numerical estimates we will use the values (\ref{ts}),
but we will also consider ``pure'' $t-J$ model which corresponds to
$t^{\prime}=t^{\prime \prime}=0$.

The rest of the paper is organized as follows. In Sec. II we describe
magnon Green's function at zero doping, and remind ideas of the 
Brueckner technique used in the work.
In Sec III the single-hole dispersion and wave function are considered.
This section is important for both Green's Function Method and for
Direct Energy Method. Sections IV and V contain the main results of the paper:
In Sec. IV we demonstrate stabilization of the dimerization
by doping using Green's Function Method, and in Sec. V we come to the
same conclusion using Direct Energy Method.
Sec. VI addresses the quantum phase transition from
the dimerized liquid to the Normal Fermi liquid at high doping.
In Sec VII we discuss shape of the Fermi surface and
distribution of the photoemission intensity.
Sec. VIII summarizes the work.
Some technical details concerning so called ``triple'' diagrams
are discussed in Appendix.

\section{Zero Doping}
At half filling ($\langle n_i\rangle =1$) the Hamiltonian (\ref{H})
is equivalent to a Heisenberg model which
has already been studied\cite{Gel,Kat}: for
 $\delta > \delta_c \approx 0.303$ the ground state is a quantum
state with gapped spectrum, and for  $\delta < \delta_c$ there is
spontaneous Neel ordering with gapless spin waves.

In order to study the stability of the dimer phase we first derive an effective
Hamiltonian in terms of bosonic operators creating spin-wave triplets (magnons)
$t^{\dag}_{i\alpha}$, $\alpha=x,y,z$ and fermionic operators creating holes 
$b^{\dag}_{\sigma}$, $a^{\dag}_{\sigma}$, $\sigma =\uparrow, \downarrow$
from the spin singlets 
shown in Fig. 1.   
This Hamiltonian consists of four parts: the spin-wave part $H_t$,
the hole part $H_h$, the spin-wave-hole interaction $H_{th}$, and 
the hole-hole interaction $H_{hh}$. Let us start from $H_t$.
Similar effective theories have been
  derived in Refs.\cite{boson} and we only present the result:
\begin{eqnarray}
\label{Ht}
H_t&=& H_{2} + H_{3} + H_{4} + H_{U},\\
H_{2}&=&\sum_{\bf{k}, \alpha} \left\{ A_{\bf{k}}
t_{\bf{k}\alpha}^{\dagger}t_{\bf{k}\alpha} +
\frac{B_{\bf{k}}}{2}\left(t_{\bf{k}\alpha}^{\dagger}
t_{\bf{-k}\alpha}^{\dagger} + \mbox{h.c.}\right) \right \}\nonumber\\,
H_{3}&=&\sum_{1+2=3} \mbox{R}({\bf k_{1}},{\bf k_{2}})
\epsilon_{\alpha\beta\gamma} t_{\bf{k_{1}}\alpha}^{\dagger}
t_{\bf{k_{2}}\beta}^{\dagger} t_{\bf{k_{3}}\gamma} + \mbox{h.c.}\nonumber\\
H_{4}&=&\sum_{1+2=3+4} \mbox{T}({\bf k_{1}}-{\bf k_{3}})
 (\delta_{\alpha\delta}\delta_{\beta\gamma}-
\delta_{\alpha\beta}\delta_{\gamma\delta})
 t_{\bf{k_{1}}\alpha}^{\dagger}
t_{\bf{k_{2}}\beta}^{\dagger}t_{\bf{k_{3}}\gamma}
t_{\bf{k_{4}}\delta}.\nonumber
\end{eqnarray}
 We also introduce an infinite repulsion on each site, in order to
 enforce the kinematic constraint 
$t_{i \alpha}^{\dag} t_{i \beta}^{\dag} = 0$.
\begin{equation}
\label{hU} 
H_{U} = U \sum_{i,\alpha \beta} t_{i \alpha}^{\dagger}t_{i \beta}^{\dagger}
t_{i \beta}t_{i \alpha}, \ \ U \rightarrow \infty
\end{equation}
The following definitions are used in (\ref{Ht}):
\begin{eqnarray}
\label{AB}
&&A_{\bf{k}}=J_{\perp} + B_{\bf k},\\
&&B_{\bf{k}}=j(\cos{k_{y}}-0.5\cos{k_{x}}),\nonumber\\
&&\mbox{T}({\bf k})= j(0.25 \cos{k_{x}} +0.5 \cos{k_{y}}),\nonumber\\
&&\mbox{R}({\bf p},{\bf q})=0.25j(\sin{q_{x}} -\sin{p_{x}}).\nonumber
\end{eqnarray}
Throughout the paper we work in the Brillouin zone of the dimerized
 lattice.
 
At zero doping ($\langle n_i\rangle =1$)  $H_t$ is
an exact mapping of the original Hamiltonian (\ref{H}).
To analyze this case it is enough to apply the Brueckner technique 
\cite{us,Kot}.
The result for the normal spin-wave Green's function reads:
\begin{equation}
\label{Gf}
G_{N}({\bf{k}},\omega) = {{\omega + \tilde{A}_{\bf{k}}(-\omega)}\over
{ \{ \omega + \tilde{A}_{\bf{k}}(-\omega) \} \{ \omega -
\tilde{A}_{\bf{k}}(\omega) \} +\tilde B_{\bf{k}}^{2}}}
\end{equation}
where 
\begin{eqnarray}
\label{ABT}
&&\tilde{A}_{\bf{k}}(\omega)= A_{\bf{k}} +\Sigma^N_4({\bf k})
+\Sigma_{Br}^{(1)}({\bf{k}},\omega),\\
&&\tilde{B}_{\bf{k}}(\omega)= B_{\bf{k}} +\Sigma^A_4({\bf k}).\nonumber
\end{eqnarray}
Normal $\Sigma_4^N$ and anomalous $\Sigma_4^A$ self-energies  are 
caused by the quartic 
interaction $H_4$ and the most important contribution
 $\Sigma_{Br}^{(1)}$ comes from the Brueckner diagrams as described in 
\cite{us}.
Strictly speaking there is also some contribution to the self-energy
caused by the "triple" interaction $H_3$. However this contribution
is very small (see, e.g. Ref.\cite{Kot}) and therefore we neglect it.

Expansion of the self-energy in powers of $\omega$ near $\omega=0$
gives quasiparticle residue and  spin-wave spectrum
\begin{eqnarray}
\label{Zo}
&&Z_{\bf{k}} = 
\left(1 - \frac{\partial \Sigma_{Br}^{(1)}}{\partial\omega}\right)^{-1},\\
&&\omega_{\bf{k}} = Z_{\bf{k}} \sqrt{[ \tilde{A_{\bf{k}}}(0)]^{2}
- \tilde{B_{\bf{k}}}^{2}}.\nonumber
\end{eqnarray}
Expressions for effective Bogoliubov parameters $u_{\bf k}$ and
$v_{\bf k}$ are given in \cite{us}.
The spin-wave gap $\Delta=\omega_{\bf k_0}$, ${\bf k_0}=(0,\pi)$,
obtained as a result of a selfconsistent
solution of Dyson's equations is plotted in Fig.2 (line at $x=0$). 
The critical value of the explicit dimerization (point where the gap vanishes)
$\delta_c=0.298$ is in agreement with results of series
expansions \cite{Gel} and quantum Monte Carlo simulations \cite{Kat}.
The validity of the Brueckner approximation is justified by the
smallness of the gas parameter 
$n_t=\sum_{\alpha}\langle t^{\dag}_{i\alpha}t_{i\alpha}\rangle$.
At the critical point 
$n_t=0.13$.

\begin{figure}[h]
\vspace{-2pt}
\hspace{-35pt}
\epsfxsize=10cm
\centering\leavevmode\epsfbox{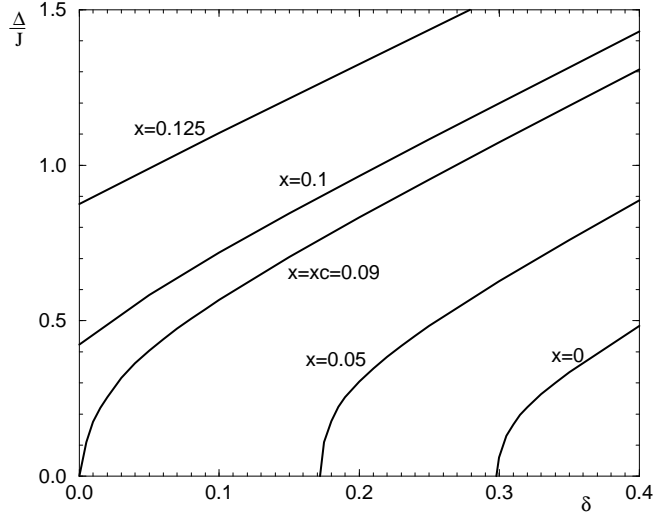}
\vspace{8pt}
\caption{\it {The magnon "gap" (centre of gravity of spectral function) as a 
function of explicit dimerization $\delta$
for $t/J=3$ and different values of hole concentration $x$,
$x=1-\langle n_i\rangle$.}}
\label{Fig2}
\end{figure}

\section{Single Hole Dispersion}
Consider now doping by holes. On the single dimer $|s\rangle$ the hole can 
exist in symmetric (bonding) and antisymmetric (antibonding) states.
Corresponding fermionic operators $b^{\dag}_{i\sigma}$, and  
$a^{\dag}_{i\sigma}$ $\sigma =\uparrow, \downarrow$  creating hole from
the singlet $|s\rangle_i$ are defined as:
\begin{eqnarray}
\label{ab}
b^{\dag}_{\sigma}|s\rangle&=&
{1\over{\sqrt{2}}}(c^{\dag}_{2,\sigma}+ c^{\dag}_{1,\sigma})|0\rangle,\\
a^{\dag}_{\sigma}|s\rangle&=&
{1\over{\sqrt{2}}}(c^{\dag}_{2,\sigma}- c^{\dag}_{1,\sigma})|0\rangle,
\nonumber
\end{eqnarray}
where $1$ and $2$ numerate the dimer sites. 

We will see later that doping suppresses
spin quantum fluctuations of the dimerized state so that
$n_t=\sum_{\alpha}\langle t^{\dag}_{i\alpha}t_{i\alpha}\rangle$
at $x > x_c$ does not exceed 0.07. This is why we can neglect these
fluctuations and consider pure dimerized spin liquid.
For comparison we can say that even for spin ladder influence of the quantum 
fluctuations on the hole dispersion is not strong \cite{Eder,sus} in
spite of the fact that in this case $n_t\approx 0.3$.

In leading approximation the wave function of a hole with given quasimomentum
is of the form
\begin{equation}
\label{1}
|1\rangle={1\over{\sqrt{N_2}}}\sum_n e^{i{\bf k}{\bf r_n}}b^{\dag}_{n\uparrow}
|S\rangle ,
\end{equation}
where $|S\rangle$ is the spin dimerized state, index $n$ numerates the dimers
and $N_2=N/2$ is number of sites in the dimerized lattice. We remind that
throughout the paper we work in the Brillouin zone of the dimerized lattice.
Sometimes the  results are transfered to the usual lattice, but then
it is specially pointed. The dispersion corresponding to the state (\ref{1})
can be easily calculated considering all possible hoppings between the dimers.
The result is
\begin{equation}
\label{e1}
\epsilon_1({\bf k})=\langle 1|H|1\rangle=
-t+(t+t')\cos k_y +({1\over 2}t+t'')\cos k_x
+t' \cos k_x \cos k_y +t''\cos 2k_y.
\end{equation}
There is also an additional t-independent constant in the dispersion
\begin{equation}
\label{e0}
e_0=1.75 J.
\end{equation}
The constant arises because a hole destroys one spin dimer
(this gives $0.75 J$) and four links $1/4n_in_j$
(see Hamiltonian (\ref{H})). In the present section we ignore $e_0$
because it gives just a shift of the entire dispersion.
However in Section V, calculating total energy of the system we will
restore $e_0$.

The wave function (\ref{1}) as well as the dispersion (\ref{e1}) 
are renormalized due to virtual admixture of antibonding states and
magnon excitations. Now we calculate this admixture to show that
it is small. In this calculation we follow the approach
developed earlier  for the doped spin ladder \cite{sus}. 
The Hamiltonian (\ref{H})
admixes following states to the wave function (\ref{1})
\begin{eqnarray}
\label{18}
|2\rangle&=&{1\over{\sqrt{N_2}}}\sum_n e^{i{\bf k}{\bf r_n}}
a^{\dag}_{n\uparrow}|S\rangle, \nonumber\\
|3\rangle&=&{1\over{\sqrt{N_2}}}\sum_n e^{i{\bf k}{\bf r_n}}
(t_n^{\dag}a^{\dag}_{n+y})|S\rangle,\nonumber \\
|4\rangle&=&{1\over{\sqrt{N_2}}}\sum_n e^{i{\bf k}{\bf r_n}}
(t_n^{\dag}a^{\dag}_{n-y})|S\rangle,\nonumber \\
|5\rangle&=&{1\over{\sqrt{N_2}}}\sum_n e^{i{\bf k}{\bf r_n}}
(t_n^{\dag}a^{\dag}_{n+x})|S\rangle,\nonumber \\
|6\rangle&=&{1\over{\sqrt{N_2}}}\sum_n e^{i{\bf k}{\bf r_n}}
(t_n^{\dag}a^{\dag}_{n-x})|S\rangle,\nonumber \\
|7\rangle&=&{1\over{\sqrt{N_2}}}\sum_n e^{i{\bf k}{\bf r_n}}
(t_n^{\dag}b^{\dag}_{n+x})|S\rangle,\nonumber \\
|8\rangle&=&{1\over{\sqrt{N_2}}}\sum_n e^{i{\bf k}{\bf r_n}}
(t_n^{\dag}b^{\dag}_{n-x})|S\rangle,\nonumber \\
|9\rangle&=&{1\over{\sqrt{N_2}}}\sum_n e^{i{\bf k}{\bf r_n}}
(t_n^{\dag}b^{\dag}_{n+x+y})|S\rangle, \\
|10\rangle&=&{1\over{\sqrt{N_2}}}\sum_n e^{i{\bf k}{\bf r_n}}
(t_n^{\dag}b^{\dag}_{n+x-y})|S\rangle,\nonumber \\
|11\rangle&=&{1\over{\sqrt{N_2}}}\sum_n e^{i{\bf k}{\bf r_n}}
(t_n^{\dag}b^{\dag}_{n-x+y})|S\rangle,\nonumber \\
|12\rangle&=&{1\over{\sqrt{N_2}}}\sum_n e^{i{\bf k}{\bf r_n}}
(t_n^{\dag}b^{\dag}_{n-x-y})|S\rangle,\nonumber \\
|13\rangle&=&{1\over{\sqrt{N_2}}}\sum_n e^{i{\bf k}{\bf r_n}}
(t_n^{\dag}a^{\dag}_{n+x+y})|S\rangle,\nonumber \\
|14\rangle&=&{1\over{\sqrt{N_2}}}\sum_n e^{i{\bf k}{\bf r_n}}
(t_n^{\dag}a^{\dag}_{n+x-y})|S\rangle,\nonumber \\
|15\rangle&=&{1\over{\sqrt{N_2}}}\sum_n e^{i{\bf k}{\bf r_n}}
(t_n^{\dag}a^{\dag}_{n-x+y})|S\rangle,\nonumber \\
|16\rangle&=&{1\over{\sqrt{N_2}}}\sum_n e^{i{\bf k}{\bf r_n}}
(t_n^{\dag}a^{\dag}_{n-x-y})|S\rangle,\nonumber \\
|17\rangle&=&{1\over{\sqrt{N_2}}}\sum_n e^{i{\bf k}{\bf r_n}}
(t_n^{\dag}b^{\dag}_{n+y})|S\rangle,\nonumber \\
|18\rangle&=&{1\over{\sqrt{N_2}}}\sum_n e^{i{\bf k}{\bf r_n}}
(t_n^{\dag}b^{\dag}_{n-y})|S\rangle.\nonumber
\end{eqnarray}
The state $|2\rangle$ is similar to $|1\rangle$ with replacement
of bonding orbital to the antibonding one. The states 
$|3\rangle - |18\rangle$ describe excited triplet (magnon) on the dimer $n$ and a 
hole on the one of the closest dimers, for example $n+x$ denotes the dimer 
on the right, $n+x+y$ denotes the up-right dimer, etc.
The brackets $(t^{\dag}a^{\dag})$ denote that spins of the magnon and the 
hole are combined to the total spin $1/2$ and z-projection $1/2$:
$|1/2,1/2\rangle$. Calculation of matrix elements of the Hamiltonian
is straightforward. The diagonal matrix elements are
\begin{eqnarray}
\label{es}
\langle 2|H|2\rangle&=&
t+(t-t')\cos k_y -({1\over 2}t-t'')\cos k_x
-t' \cos k_x \cos k_y +t''\cos 2k_y,\nonumber\\
\langle 3|H|3\rangle&=&\langle 4|H|4\rangle=t+J_{\perp}-j/2,\nonumber\\
\langle 5|H|5\rangle&=&\langle 6|H|6\rangle=t+J_{\perp}-j/4,\nonumber\\
\langle 7|H|7\rangle&=&\langle 8|H|8\rangle=-t+J_{\perp}-j/4,\\
\langle 9|H|9\rangle&=&\langle 10|H|10\rangle=
\langle 11|H|11\rangle=\langle 12|H|12\rangle=-t+J_{\perp},\nonumber\\
\langle 13|H|13\rangle&=&\langle 14|H|14\rangle=
\langle 15|H|15\rangle=\langle 16|H|16\rangle=t+J_{\perp},\nonumber\\
\langle 17|H|17\rangle&=&\langle 18|H|18\rangle=-t+J_{\perp}-j/2.\nonumber
\end{eqnarray}
The nonzero nondiagonal matrix elements are
\begin{eqnarray}
\label{nd}
\langle 2|H|1\rangle&=&-i{t\over 2}\sin k_x,\nonumber\\
\langle 3|H|1\rangle&=&\langle 4|H|1\rangle^*=
{{\sqrt{3}}\over{2}}(t-t')+{{\sqrt{3}}\over{4}} j e^{i k_y},\nonumber\\
\langle 5|H|1\rangle&=&\langle 6|H|1\rangle^*=
-{{\sqrt{3}}\over{4}}(t-2t'')-{{\sqrt{3}}\over{8}} j e^{i k_x},\nonumber\\
\langle 7|H|1\rangle&=&-\langle 8|H|1\rangle^*=
-{{\sqrt{3}}\over{4}}t+{{\sqrt{3}}\over{8}} j e^{i k_x},\nonumber\\
\langle 9|H|1\rangle&=&\langle 10|H|1\rangle=
-\langle 11|H|1\rangle=-\langle 12|H|1\rangle=-{{\sqrt{3}}\over{4}}t',
\nonumber\\
\langle 13|H|1\rangle&=&\langle 14|H|1\rangle=
\langle 15|H|1\rangle=\langle 16|H|1\rangle=-{{\sqrt{3}}\over{4}}t',\\
\langle 5|H|2\rangle&=&-\langle 6|H|2\rangle^*=
{{\sqrt{3}}\over{4}}t+{{\sqrt{3}}\over{8}} j e^{i k_x},\nonumber\\
\langle 7|H|2\rangle&=&\langle 8|H|2\rangle^*=
{{\sqrt{3}}\over{4}}(t+t'')-{{\sqrt{3}}\over{8}} j e^{i k_x},\nonumber\\
\langle 9|H|2\rangle&=&\langle 10|H|2\rangle=
\langle 11|H|2\rangle=\langle 12|H|2\rangle={{\sqrt{3}}\over{4}}t',
\nonumber\\
\langle 13|H|2\rangle&=&\langle 14|H|2\rangle=
-\langle 15|H|2\rangle=-\langle 16|H|2\rangle={{\sqrt{3}}\over{4}}t',
\nonumber\\
\langle 17|H|2\rangle&=&\langle 18|H|2\rangle^*=
{{\sqrt{3}}\over{2}}(t+t')+{{\sqrt{3}}\over{4}} j e^{i k_y},\nonumber\\
\langle 4|H|3\rangle&=&-{{(t-t')}\over{2}} e^{-i k_y},\nonumber\\
\langle 18|H|17\rangle&=&-{{(t+t')}\over{2}} e^{-i k_y}.\nonumber
\end{eqnarray}
Diagonalization of the Hamiltonian matrix can be performed numerically. 
This gives the quasiparticle dispersion $\epsilon_{\bf k}$
and the quasiparticle residue $Z_{\bf k}^{(h)}$  which by definition is 
equal to the weight of the state $|1\rangle$ (see eq. (\ref{1}) in the exact 
wave function \cite{com}.
The minimum energy for different sets of parameters is shown in the fifth column of the
Table I. We remind that we use units $J=1$. In the last column we show position of the
minimum. For comparison we also display the minimum value of $\epsilon_1$ (see eq. (\ref{e1})),
which always is at ${\bf p}_0=(\pi,\pi)$.
\vspace{10pt}
\begin{center}
Table I. \hspace{1.cm}
\begin{tabular}{|c|c|c|c|c|c|}\hline
$t \ \ $ & $t'$ \ \ & $t'' \ \ $ & $\epsilon_{1,min}$ \ \ & $\epsilon_{min}$ \ \ & 
${\bf p}_0$ \ \ \\ \hline
3.1 & -0.8 & 0.7 & -7.75 & -9.70 & $(\pi,0.82\pi)$ \\ \hline
3. & 0. & 0. & -7.50 & -9.03 & $(\pi,\pi)$ \\ \hline
2. & 0. & 0. & -5.0 & -5.95 & $(\pi,\pi)$ \\ \hline
1. & 0. & 0. & -2.5 & -2.91 & $(\pi,\pi)$ \\ \hline
\end{tabular}
\vspace{10pt}
\end{center}
Let us denote the hole concentration by $x=n/N$, where $n$ is  number of holes,
and $N$ is number of sites. Hence on-site electron occupation number is
$\langle n_i\rangle=1-x$.
Concentration of holes in  terms of the dimerized lattice is two times larger
$n/(0.5N)=2x$.
At this stage we neglect interaction of holes between themselves, hence we
consider them as an ideal Fermi gas, and the Fermi surface can be easily found
from the condition
\begin{equation}
\label{cons}
2\int {{dk_x dk_y}\over{(2\pi)^2}}=2x,
\end{equation}
where integration is performed inside the Brillouin zone of the dimerized
lattice.  The Fermi surface at $x=0.1$ and hopping parameters
given in (\ref{ts}) is shown in Fig. 3 by solid line.

\begin{figure}[h]
\vspace{20pt}
\hspace{-35pt}
\epsfxsize=10cm
\centering\leavevmode\epsfbox{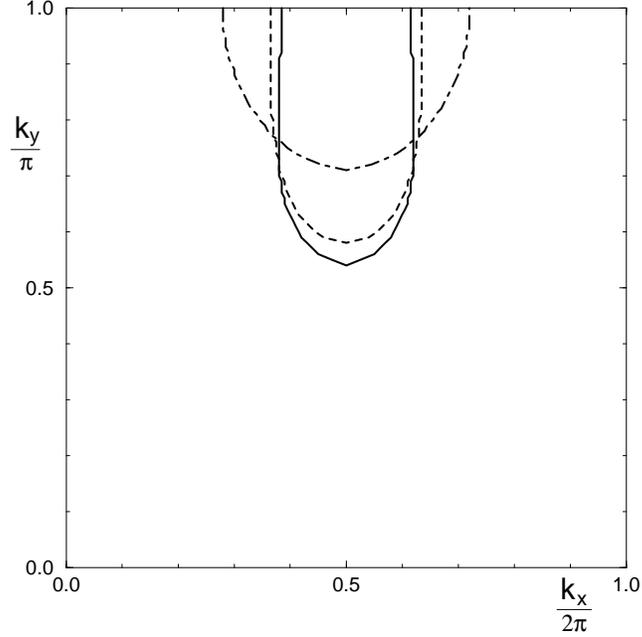}
\vspace{4pt}
\caption{\it {Fermi surface of the dimerized spin liquid at doping $x=0.1$.
Dot-dashed line corresponds to $t/J=3$, $t',t''=0$,
solid and dashed lines correspond to the set of parameters from
(\ref{ts}). Solid line gives Fermi surface found with ``exact''
dispersion as it is described in Section III. Dashed line
corresponds to the bare dispersion (\ref{e1}).
We put $k_x/2\pi$ along the horizontal axis and $k_y/\pi$ along the 
vertical axis, so  the picture corresponds to a quadrant of
the Brillouin zone of the original lattice.}}
\label{Fig3}
\end{figure}

We stress that in this 
figure we put $k_x/2\pi$ along the horizontal axis and $k_y/\pi$ along the 
vertical axis. It means that the picture corresponds to a quadrant of
the Brillouin zone of the original lattice. 
The Fermi surface corresponding
to the bare dispersion 
(\ref{e1}) is shown by the dashed line. The two curves are very close and
this proves that the dispersion renormalization is small. The 
wave function renormalization is also not large and the 
quasiparticle residue is close to unity: it is
$Z^{(h)}=0.83$ at the bottom
of the band, and at the Fermi surface $Z^{(h)}=0.80$.
So admixture of the states (\ref{18}) to the bare state $|1\rangle$
is relatively small. In this admixture the states $|3\rangle$, $|4\rangle$, 
$|7\rangle$, and $|8\rangle$ clearly dominate. In a reasonable
approximation the wave function can be written as
\begin{equation}
\label{crude}
\psi \approx 0.9|1\rangle-0.22(|3\rangle+|4\rangle-|7\rangle+|8\rangle).
\end{equation}
Admixture of other components is even smaller.
For comparison in Fig. 3 we show also by dot-dashed line the Fermi surface 
for ``pure'' $t-J$ model ($t/J=3$, $t'=t''=0$, doping is the same, $x=0.1$). 
We see that the additional  hoppings influence substantially shape of the 
Fermi  surface.

\section{The Spin-Wave-Hole Interaction. Stabilization of the 
Dimer Order}

The magnon-hole interaction $H_{th}$ can be easily calculated in the way
similar to that for doped spin-ladder \cite{Eder,sus}.
This interaction consists of two parts. The first one is interaction of a
hole and a magnon positioned at different dimers. This is a relatively weak
interaction which can be neglected \cite{Eder,sus}. The second part, which 
gives the main effect, comes from the
constraint that a hole and a magnon can not coexist at
the same dimer: 
$t^{\dag}_{i\alpha}b^{\dag}_{i\sigma}=t^{\dag}_{i\alpha}a^{\dag}_{i\sigma}=0$. 
To deal with this constraint we
introduce, similarly to (\ref{hU}), an infinite repulsion
\begin{equation}
\label{hU1} 
H_{U1} = U \sum_{i,\alpha \sigma} t_{i\alpha}^{\dagger}t_{i\alpha}
(b_{i\sigma}^{\dagger}b_{i\sigma}+a_{i\sigma}^{\dagger}a_{i\sigma})
, \ \ U \rightarrow \infty.
\end{equation}
The exact hole-magnon scattering amplitude caused by this
interaction can be found via Bethe-Salpeter
equation shown in Fig.4a. 

\begin{figure}[h]
\vspace{-2pt}
\hspace{-35pt}
\epsfxsize=15cm
\centering\leavevmode\epsfbox{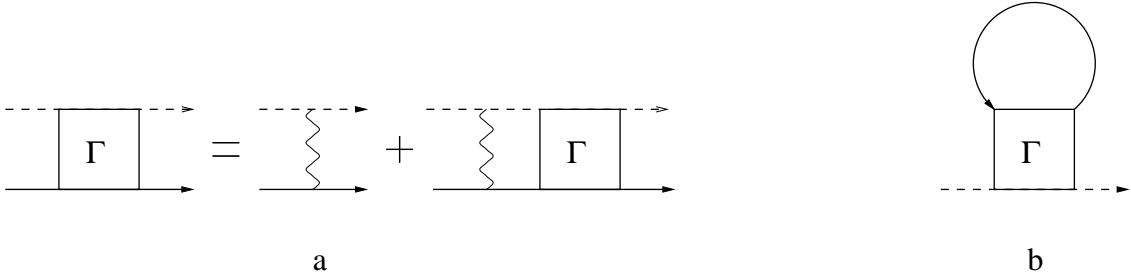}
\vspace{8pt}
\caption{\it {
(a) Bethe-Salpeter equation for hole-magnon scattering vertex $\Gamma$.
Solid line corresponds to the hole and dashed line to the magnon.
(b) Brueckner contribution to the magnon self energy.}}
\label{Fig4}
\end{figure}

This scattering amplitude is similar to that for magnon-magnon 
scattering \cite{us}. Solution of the Bethe-Salpeter equation gives 
\begin{equation}
\label{gam}
\Gamma(E,{\bf k})=-\left(\sum_{\bf q} {{Z_{\bf q}u_{\bf q}^2}\over
{E-\omega_{\bf q}-\epsilon_1({\bf k-q})}}\right)^{-1},
\end{equation}
where $E$ and ${\bf k}$ is total energy and total momentum of the incoming
particles.
It has been demonstrated in the previous section that the renormalized
hole dispersion $\epsilon_{\bf k}$ is close to the bare one
$\epsilon_1({\bf k})$. This is why we use  in eq. (\ref{gam}) the
bare dispersion (\ref{e1}).

We remind that concentration of holes in  terms of the dimerized lattice is 
$2x \ll 1$, and this is the gas parameter of the
magnon-hole Brueckner approximation. In the previous section we have shown that
the holes are concentrated in the pocket in the vicinity of 
${\bf p_0}=(\pi,\pi)$. Therefore the magnon
normal self-energy described by the diagram Fig. 4b is
\begin{equation}
\label{sh}
\Sigma_{Br}^{(2)}({\bf k},\omega)=
2x \Gamma [\omega+\epsilon_1({\bf p_0}),{\bf k+p_0}]
\end{equation}

It is instructive to consider first the limit which allows an 
analytical solution: $j \ll J_{\perp}$, $\sqrt{2}\pi x \ll 1$. 
Bare magnon dispersion in this case is 
$\omega_{\bf k}\approx J_{\perp}+j(\cos k_y-0.5\cos k_x)$ and hence the 
integrals in (\ref{gam},\ref{sh}) can be calculated analytically with 
logarithmic accuracy. This gives
\begin{equation}
\label{spt}
\Sigma_{Br}^{(2)}({\bf k},\omega)\approx {{2\sqrt{2}\pi x (t+j)}\over
{\ln(12.5/\mu) +i\pi \theta(\delta\omega)}},
\end{equation}
where 
\begin{eqnarray}
\label{dm}
&&\delta\omega = {1\over{t+j}}\left[\omega-\omega_{\bf k}+
{j\over{t+j}}(\omega_{\bf k}-\omega_{\bf k_0})\right],\\
&&\mu= \max(|\delta\omega|,\sqrt{2}\pi x),\nonumber
\end{eqnarray}
and $\theta(\delta\omega)$ is a step function.
The magnon Green's function is
\begin{equation}
\label{gf}
G({\bf k},\omega)=
{1\over{\omega-\omega_{\bf k}-\Sigma_{Br}^{(2)}({\bf k},\omega)}}.
\end{equation}
For illustration the spectral function $Im G(\omega)$
at  ${\bf k}={\bf k_0}=(0,\pi)$, $t/j=3$ and different $x$ is plotted in Fig.5.

\begin{figure}[h]
\vspace{-10pt}
\hspace{-35pt}
\epsfxsize=10cm
\centering\leavevmode\epsfbox{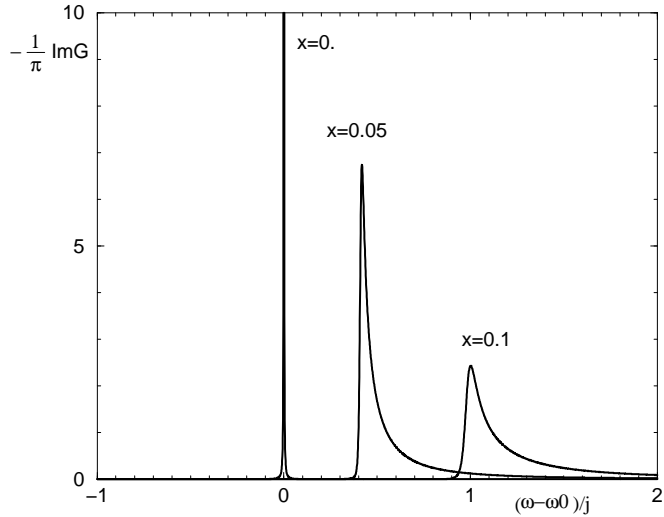}
\vspace{-20pt}
\caption{\it {Magnon spectral density at ${\bf k}={\bf k_0}=(0,\pi)$ in the limit 
$J_{\perp}\gg j$, $t/j=3$, and different hole concentrations.}}
\label{Fig5}
\end{figure}

There are several conclusions from formula (\ref{spt}) and Fig. 5:
1) doping pushes the spin-wave spectrum up, 2) the effect is increasing with
hopping $t$, 3) finite width appears, 4) there is only 
a logarithmic dependence on the infrared cutoff.
Let us stress the importance of the point (4). It means that the
effect is practically independent of the long-range dynamics.
 Moreover, near the critical point ($\Delta=0$) the
 situation is even better: the spin-wave spectrum is linear
 and even the logarithmic divergence disappears.
Thus in the 2D case there is separation of scales which
justifies Brueckner approximation. If we 
tried to apply the described approach to the 1D case (say a doped spin ladder)
we would get into trouble: power infrared divergence appears in Brueckner 
diagram and  hence there is no justification for gas approximation.
Let us also comment on the point (3) (width). There is also a
"triple" contribution  to the magnon self-energy shown in  Fig. 6a,b.

\begin{figure}[h]
\vspace{-2pt}
\hspace{-35pt}
\epsfxsize=10cm
\centering\leavevmode\epsfbox{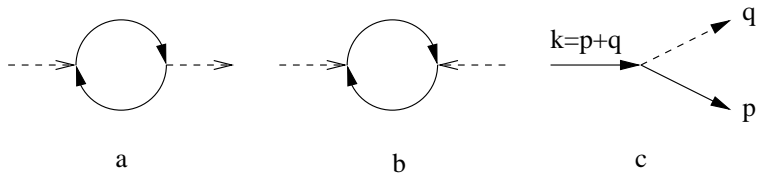}
\vspace{10pt}
\caption{\it {
(a) Normal ``triple'' contribution to the magnon self energy.
(b) Anomalous ``triple'' contribution to the magnon self energy.
(c) hole-magnon ``triple'' vertex.}}
\label{Fig6}
\end{figure}

This is a long-range contribution, and 
it can be shown that it does not influence position of the critical point 
($\Delta=0$) in linear in x approximation. This fact will be proved
in Appendix and this is why here we neglect the ``triple''
diagrams. However note that these diagrams influence the width of the magnon 
spectral function.

In the general case there are two contributions to the Brueckner self-energy:
$\Sigma_{Br}^{(1)}$, which is due to the magnon-magnon constraint, and
$\Sigma_{Br}^{(2)}$ which is due to the magnon-hole constraint.
To find the spin wave spectrum one has to solve selfconsistently
Dyson's equation for Green's function (\ref{Gf}), as it is described in Ref.
\cite{us}.  Results for the spin-wave "gap" $\Delta$  as a function of 
explicit dimerization $\delta$ for different hole concentrations $x$ and 
$t/J=3$ are plotted in Fig. 2. These curves are practically independent
of the longer range hoppings $t'$ and $t''$. Strictly speaking
at $x \ne 0$ the $\Delta$ is not a gap because of the large decay width.
What we plot is the position of the center of gravity of the magnon spectral
function.  However at  $\Delta \to 0$ the width vanishes, and therefore the 
critical regime is uniquely defined. 

It is clear from Fig.2 that at $t/J=3$ and $x > x_{c1} \approx 0.090$ the 
"gap"  remains finite even at $\delta =0$.
This is regime of spontaneous dimerization.
Critical concentrations for other values of $t/J$ are presented in Table II.
\vspace{10pt}
\begin{center}
Table II. \hspace{1.cm}
\begin{tabular}{|c|c|c|c|}\hline
$t/J$ \ \ \  & \ \ \ 1 \ \ \  & \ \ \ 2 \ \ \  & \ \ \ 3 \\
&&&\\
 $x_{c1}$ & 0.132 & 0.106 & 0.090\\ \hline
\end{tabular}
\vspace{10pt}
\end{center}
 Thus the doping stabilizes the dimerized phase. 
The larger the hopping $t$, the stronger the effect of stabilization.
(The same follows from  eq. (\ref{spt}).)
This statement is true only if $t/J \lesssim 10$. At $t/J \sim 10$
there is a crossover to quasiparticles with higher spin (hole-magnon 
bound states) \cite{sus} which indicates transition to
the Nagaoka regime.
The small parameter of the Brueckner approximation is concentration of 
holes in the dimerized lattice: $2x$. Therefore at
$t/J=3$ one should expect $\sim 20\%$  accuracy in calculation of $x_{c1}$.
Note that the value of $x_{c1}$ is close to that found in \cite{Ig} from the
Neel state.

Another important gas parameter is density
of spin fluctuations $n_t$. It also proved to be small:
At the critical point, 
$\delta=0$, $x=x_c=0.09$, 
the density is $n_t\approx 0.07$.

\section{Direct Comparison of Ground State Energies of the Neel State
and the Spin Dimerized State}

In the previous section we have demonstrated quantum phase transition
from the Neel state to the spin dimerized state at some critical hole
concentration $x_{c1}$. We calculated the magnon Green's function
in the dimerized phase, and the transition point was identified as the
point where the magnon gap vanishes. This is a rigorous approach
which allows to determine $x_{c1}$ with relatively high precision
($\sim 20\%$), however it is rather involved
technically. An alternative method is direct comparison of the ground state
energies of these two states. This is a very simple method which does
not require introduction of the explicit dimerization. So throughout
this section $\delta=0$.

The energy per site for the undoped 
Neel state is $-1.17=-0.67-0.5$, where the first term is Heisenberg energy
(see e. g. Ref. \cite{Man}), and the second contribution comes from  the
$-1/4n_in_j$ term in the Hamiltonian (\ref{H}). 
We remind that in our units J=1.
In this section we consider only ``pure'' $t-J$ model, i. e. $t'=t''=0$.
Energy of a single hole injected into the Neel background is 
$-3.17t+2.83 t^{0.27}+1$, see e. g. Ref.\cite{Dag}, where the last term
comes from the $-1/4n_in_j$ term in the Hamiltonian (\ref{H}).
Therefore in linear in x approximation energy of the doped Neel state is
\begin{equation}
\label{EN}
E_{Neel}/N=-1.17+(-3.17t+2.83 t^{0.27}+1)x.
\end{equation}

Energy of the undoped columnar dimerized state without account of quantum 
fluctuations is $(-0.375-0.5)N$, where $-0.375$ is Heisenberg energy and $-0.5$
comes from the $-1/4n_in_j$ term in the Hamiltonian (\ref{H}).
Quantum fluctuations push this energy down. Using perturbation
theory one can find that in linear in $n_t$ (triplet density)
approximation this shift is $\sim -0.5n_t N$. According to previous section
at the critical point $n_t \approx 0.07$ and therefore the shift is tiny.
Energy of a single hole injected into the spin dimerized background
has been found in section III, $\epsilon=e_0+\epsilon_{min}$, where $e_0$ is
given by eq. (\ref{e0}) and the values of $\epsilon_{min}$ are
presented in Table I. For $1\le t \le 4$ one can fit $\epsilon_{min}$
as $\epsilon_{min}\approx-3.0t$. Altogether this gives following energy of the
doped dimerized state (linear in x approximation)
\begin{equation}
\label{ED}
E_{Dimer}/N=-0.91+(-3t+1.75)x.
\end{equation}

At $x=0$ the Neel state energy (\ref{EN}) is lower than that of the
dimerized state (\ref{ED}). The critical concentration (the transition 
point to the dimerized state) is defined by the condition 
$E_{Neel}=E_{Dimer}$. This gives following values of the critical 
concentration:
\vspace{10pt}

\begin{center}
Table III. \hspace{1.cm}
\begin{tabular}{|c|c|c|c|}\hline
$t/J$ \ \ \  & \ \ \ 1 \ \ \  & \ \ \ 2 \ \ \  & \ \ \ 3 \\
&&&\\
 $x_{c1}$ & 0.14 & 0.11 & 0.10\\ \hline
\end{tabular}
\vspace{10pt}
\end{center}

The values of $x_{c1}$ in Table III are somewhat overestimated.
The matter is that the single hole energy for the Neel state
is known with high precision, at the same time
similar energy for the dimerized state has been found in Section III
with trial wave function which contains only 18 components. 
The true energy is lower than the variational one. To estimate this uncertainty
we refer to the doped spin ladder. There is a variational calculation \cite{sus} 
for this system which is similar to to the calculation in Sec. III, and there 
are also exact numerical simulations \cite{Troyer}.
Comparison shows that the variational method underestimate $\epsilon_{min}$
(see Table I) by 10-15\%. Taking this as an estimate we should replace
the term $-3t$ in eq. (\ref{ED}) by $-3.3t$. Then we come to following
values of the critical concentration
\vspace{10pt}

\begin{center}
Table IV. \hspace{1.cm}
\begin{tabular}{|c|c|c|c|}\hline
$t/J$ \ \ \  & \ \ \ 1 \ \ \  & \ \ \ 2 \ \ \  & \ \ \ 3 \\
&&&\\
 $x_{c1}$ & 0.12 & 0.09 & 0.08\\ \hline
\end{tabular}
\vspace{10pt}
\end{center}

Altogether results for $x_{c1}$ presented in Tables II, and III, IV
and derived by absolutely different methods are in remarkable agreement. 
This gives a very strong confirmation of the phase transition to the 
columnar spin dimerized state. Physical reason for stability of the spin
dimerized state is especially evident after the energy considerations:
this is the gain in the hole kinetic energy, it is ``easier'' to propagate 
in the dimerized background.

\section{Spin-Dimer Order Parameter and Transition to the Normal
Fermi Liquid at High Doping.}

We have discussed the transition to the Neel state at
 hole concentration $x< x_{c1}$.
It is clear that at large $x$ ($x > x_{c2}$) there is a 2nd order phase
transition to the normal  Fermi liquid. 
Let us define the spin-dimer order parameter as
\begin{equation}
\label{ord}
\rho=\langle {\bf S}_2{\bf S}_3\rangle-\langle {\bf S}_1{\bf S}_2\rangle,
\end{equation}
where sites 1, 2, and 3 are shown in Fig. 1. For perfect dimerized state
$\rho= 3/4$. There are two mechanisms for reduction of the order parameter.
The first one is due to spin quantum fluctuations which approximately give
$\rho \to 3/4-n_t$. 
The second mechanism is direct effect of doping. Naive estimate is
$\rho \to 3/4(1-2x)$, however the hole wave function (\ref{crude})
is slightly different from the bare one and because of this the coefficient
in the naive formula is slightly renormalized. All together this gives
\begin{equation}
\label{ord1}
\rho={3\over 4}[1-2x(1+6\alpha^2)]-n_t,
\end{equation}
where $\alpha=0.22$ is the admixture coefficient in eq. (\ref{crude}).
This formula is derived in dilute gas approximation, i.e at $2x, n_t \ll 1$,
however for an estimate we can extend it to large $x$. Setting
$n_t \sim0.05$ we find from (\ref{ord1}) that $\rho$ vanishes
at $x_{c2}\approx 0.36$. We repeat that this is
only an estimate because the approach  assumes that $2x \ll 1$.

The phase diagram of the $t-J-\delta$ model at
zero temperature is presented in Fig. 7

\begin{figure}[h]
\vspace{-2pt}
\hspace{-35pt}
\epsfxsize=10cm
\centering\leavevmode\epsfbox{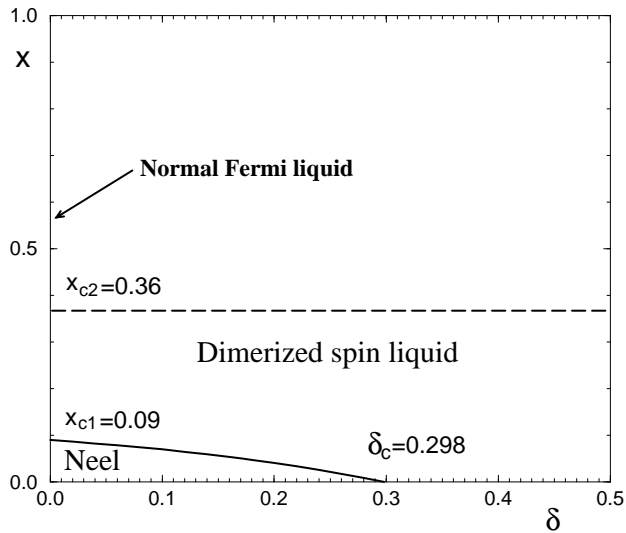}
\vspace{-10pt}
\caption{\it {Phase diagram of the $t-J-\delta$ model ($t/J=3$) in the 
plane doping ($x$) - explicit dimerization ($\delta$).}}
\label{Fig7}
\end{figure}

Because of the mobile holes the dimerized spin liquid 
at $x_{c1}<x<x_{c2}$ is a conducting state.
Stability of this state is a very robust
effect because it is due to the high energy correlations
(typical energy scale $\sim 2t$). There are also low energy
effects with typical energy scale $\sim 2tx$ which can lead
to hole-hole pairing, charge stripes, etc.
We do not consider these effects in the present work because they are 
secondary with respect to the main one: spin dimerization. 
However we would like to note that there is a simple mechanism for charge 
stripes induced by the spin dimers: Because of the anisotropic dispersion,
see Fig. 3,  the charge response is enhanced at some momentum 
${\bf p}=(p_x,0)$. The effect is very sensitive to additional hopping 
parameters $t'$ and $t''$, they
can further enhance or suppress the response. 

\section{Shape of the Fermi Surface and PES Intensity}
Shape of the Fermi surface for the dimerized state at the doping $x=0.1$
and hopping matrix elements from (\ref{ts}) is shown in Fig.3
by the solid line. In this  figure we put $k_x/2\pi$ along the horizontal 
axis and $k_y/\pi$ along the vertical axis, where ${\bf k}$ is defined
on the dimerized lattice. In terms of the Brillouin zone  of the original 
lattice this is usual quadrant: $0 \le P_x \le \pi$, $0 \le P_y \le \pi$.
To distinguish between the dimerized lattice and the original one we
denote by capital letters momenta corresponding to the original lattice:
$P_x=k_x/2$, $P_y=k_y$.

In a real sample there are domains with one stripe dimerization and there
are domains with stripes rotated by $90^o$. Therefore in an experiment
one should see two superimposed Fermi surfaces. Corresponding picture for
$x=0.15$ is shown in Fig. 8.

\begin{figure}[h]
\vspace{10pt}
\hspace{-35pt}
\epsfxsize=10cm
\centering\leavevmode\epsfbox{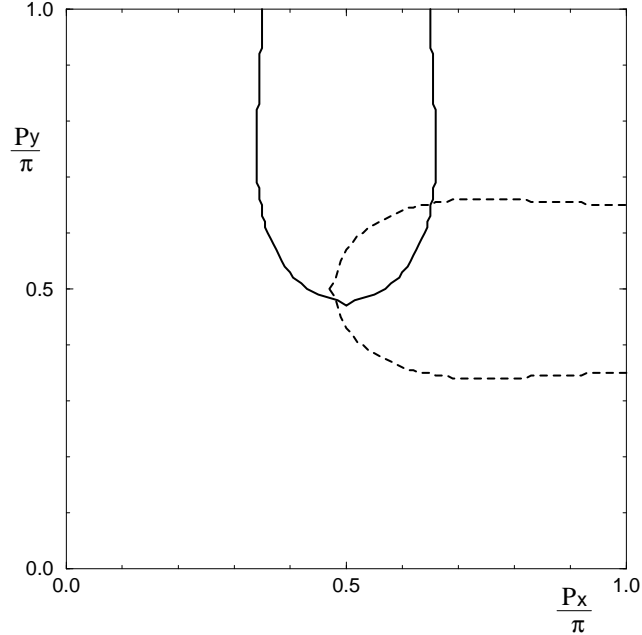}
\vspace{-10pt}
\caption{\it {Two superimposed Fermi surfaces  of the dimerized spin liquid 
corresponding to different orientations of the stripes.
The hole concentration is $x=0.15$.
The picture corresponds to a quadrant of
the Brillouin zone of the original lattice $0\le P_x \le \pi$, 
$0\le P_y \le \pi$.}}
\label{Fig8}
\end{figure}

The first impression is that it is very much different from what is
usually observed in angular resolved photoemission (PES) experiments,
see e. g. Ref. \cite{PES}
However let us calculate intensity of the photoemission.

The photoeffect operator is of the form
\begin{equation}
\label{ph}
\hat A =\sum_n c_{i\downarrow}e^{i{\bf P}{\bf r_n}},
\end{equation}
 where summation is performed over sites of the square lattice.
Amplitude of the hole creation from the dimerized background (oriented as
it is shown in Fig. 1) is equal to
\begin{equation}
\label{am}
A=\langle s| b_{\uparrow} \hat A|s\rangle=\cos (P_x/2).
\end{equation}
Here we have taken into account that according to considerations in
Section III wave function of the hole in bonding state is
practically unrenormalized. We stress once more that {\bf P} is a
quasimomentum corresponding to the original lattice.
According to (\ref{am}) intensity of PES spectra ($I \propto
A^2$) is dropping quickly as $P_x$ is increasing.
If one assumes that the $t-J$ model originates from the single band
Hubbard model, then the corrections of the order of $t/U$ to the PES intensity 
can be calculated in a way suggested in Ref. \cite{Saw}. This gives
\begin{equation}
\label{int}
I_{\bf P}\propto \left(\cos{{P_x}\over{2}}+{J\over{8t}}\cos{{3P_x}\over{2}}
+{J\over{4t}}\cos{{P_x}\over{2}} \cos P_y\right)^2.
\end{equation}
The Hubbard repulsion $U$ is excluded from this formula using
relation $J=4t^2/U$. 
According to (\ref{int}) the PES intensity is strongly asymmetric at the
Fermi surface. For example at $x=0.15$ the intensity at the right top
corner of the Fermi surface, $P_x=0.66\pi$, $P_y=\pi$ (see Fig. 8) is
3.5 times smaller than that at the left top corner of the Fermi surface,
$P_x=0.34\pi$, $P_y=\pi$. In  real cuprate the asymmetry must be
even stronger. The reason for further enhancement of the asymmetry is
structure factor of the Zhang-Rice singlet. This is quite similar to 
the well understood situation in  the charge transfer insulator 
Sr$_2$CuO$_2$Cl$_2$, see Ref. \cite{Saw}.

Thus the angle resolved photoemission measurements are sensitive
only to the ``inner'' parts (the parts closest to the $\Gamma$ point: ${\bf P}=(0,0)$)
of the Fermi surfaces shown in Fig. 8. Shape of this effective
``Fermi surface'' is very close to what is observed in numerous 
PES experiments. Another feature which agrees with experiment is
width of the ``quasiparticle'' peak along (1,1) direction:
it is always rather big because the peak arises as a superposition of
two different peaks corresponding to two different Fermi surfaces.

\section{Conclusions}

In conclusion, using the dilute gas approximation we have analyzed the 
phase diagram of the $t-J$ model and the stability of spin-dimerized 
phase. The main result of the work is phase diagram shown in Fig. 7.
Without any explicit dimerization ($\delta=0$) the spin dimerized phase
is stable at hole concentration $x_{c1}< x <x_{c2}$. At $t/J=3$ the critical
concentrations are $x_{c1}\approx 0.09$,
$x_{c2}\sim 0.36$. At $x<x_{c1}$ the system undergoes transition to the Neel 
state, and at $x>x_{c2}$ to the Normal Fermi liquid.

To prove stability of the spin dimerized phase and to calculate critical
concentrations we have used two independent approaches. The first one is 
based on the calculation of the magnon Green's function. 
The second approach consists in direct comparison of ground state
energies of the Neel state and the dimerized state.
Both approaches demonstrate stability of the spin dimerized phase and
give very close values of the critical concentrations.

\acknowledgments

I thank V. N. Kotov and M. Yu. Kuchiev for stimulating discussions.
I am especially grateful to T. M. Rice who attracted my attention to
the direct comparison of the ground state energies.

\section{Appendix: ``Triple'' Diagrams}

Purpose of the present section is to demonstrate that the ``triple''
diagrams shown in Fig. 6a,b do not influence position of the critical 
point $x_{c1}$ found in Section IV.
The ``triple'' vertex is shown in Fig. 6c. 
In the present section we completely neglect small renormalization
of the hole wave function considered in Section III. Therefore,
the initial state in Fig. 6c is given by
(\ref{1}) and the final state is
\begin{equation}
\label{f}
|f\rangle=b^{\dag}_{p\sigma}t^{\dag}_{q\alpha}|S\rangle=
{1\over{N_2}}\sum_n e^{i{\bf p}{\bf r_n}}b^{\dag}_{n\sigma}
\sum_m e^{i{\bf q}{\bf r_m}}t^{\dag}_{m\alpha}|S\rangle.
\end{equation}
Kinematic structure of the vertex is obvious
\begin{equation}
\label{v}
\Gamma_{\bf p,q}=-iA t^{\dag}_{\bf q\alpha}[b^{\dag}_{\bf p}
\sigma_{\alpha}b_{\bf p+q}],
\end{equation}
where $\sigma_{\alpha}$, $\alpha =1,2,3$ is vector of Pauli matrices, 
and $b^{\dag}_{\bf p}$ is the hole wave function in spinor representation.
Direct calculation of the matrix element $\langle f|H|1\rangle$ and
comparison with (\ref{v}) gives following value of the coupling constant
\begin{equation}
\label{A}
A={1\over 2}\left[(t+2t'\cos p_y)\sin p_x+{j\over 2}\sin q_x\right].
\end{equation}
The normal ``triple'' magnon self-energy $\Sigma_{3n}({\bf q},\omega)$
is shown in Fig. 6a.
Since we are interested in the critical point $x_{c1}$ it is enough
to calculate the self energy at zero frequency and at the momentum
where the spin-wave gap vanishes: ${\bf q}= {\bf q}_0=(0,\pi)$.
Straightforward calculation of the loop at $t'=t''=0$ gives
\begin{equation}
\label{3n}
\Sigma_{3n}({\bf q}_0,0)=-t x.
\end{equation}
Unfortunately analytical calculation at the values of $t'$ and $t''$
from (\ref{ts}) is impossible because at these values the curvature of the
hole dispersion along y-direction vanishes. However numerical calculation
shows that at $x =0.1-0.15$ and $t'$, $t''$ from (\ref{ts}) the "triple"
self-energy is by a factor 1.5 smaller than one given by (\ref{3n}).
Comparing the ``triple'' self-energy with the Brueckner one (\ref{sh}) we 
find that the "triple" self-energy is by a factor 7 smaller.
 This is already enough to neglect the "triple"
contribution. However we would like to demonstrate that suppression of
the "triple" contribution is even stronger: its influence on the point of
phase transition is exact(!) zero in linear in x approximation. This 
interesting fact is related to anomalous "triple" self energy shown in Fig. 6b. Simple consideration based 
on the structure of the vertex (\ref{v}) shows that following exact relation
takes place
\begin{equation}
\label{3a}
\Sigma_{3a}({\bf q}_0,0)=-\Sigma_{3n}({\bf q}_0,0).
\end{equation}

According to (\ref{Zo}) the magnon spectrum found  without "triple" diagram
is of the form
\begin{equation}
\label{o}
\omega_{\bf q} \propto \sqrt{{\tilde A}^2-{\tilde B}^2},
\end{equation}
where ${\tilde A}$ arises from the normal terms in the effective
Hamiltonian and ${\tilde B}$ arises from the anomalous terms.
At the critical point ${\tilde A}=-{\tilde B}$ and hence the excitation
energy vanishes. With account of "triple" diagrams the relation (\ref{o})
should be rewritten as
\begin{equation}
\label{o1}
\omega_{\bf q} \propto \sqrt{\left({\tilde A}+\Sigma_{3n}\right)^2-
\left({\tilde B}+\Sigma_{3a}\right)^2}.
\end{equation}
It is clear that because of (\ref{3a}) the  dispersion 
(\ref{o1}) vanishes exactly at the same point as (\ref{o}).
This proves our statement that the "triple" self energy does not influence
position of the transition point. However away from the transition point
when the spin-wave gap is increasing the "triple" self energy is getting more
important. Note that similar situation takes place with
"triple" diagrams considered in Ref. \cite{Kot} for $J_1-J_2$ model.

\end{document}